\documentclass[9pt,twocolumn,twoside]{osajnl}

\journal{ol} 

\setboolean{shortarticle}{false} 

\ifthenelse{\boolean{shortarticle}}{\colorlet{color2}{color2b}}{\colorlet{color2}{color2}} 

\newcommand{\sss}{\scriptscriptstyle}
\newcommand{\sst}{\scriptstyle}

\newcommand{\stext}[1]{\sss \text{#1} \sst}

\title{\vspace{-2cm}Broadband near-infrared antireflection coatings fabricated by three dimensional direct laser writing}
\author[1]{Y. Li}
\author[1]{D.B. Fullager}
\author[1]{E. Angelbello}
\author[2]{D. Childers}
\author[1]{G. Boreman}
\author[1,3,*]{T. Hofmann}

\affil[1]{Department of Physics and Optical Science, University of North Carolina at Charlotte, 9139 University City Blvd., Charlotte, NC 28223, USA}
\affil[2]{USCONEC,1138 25th Street Southeast, Hickory, NC 28602, USA}
\affil[3]{Department of Physics, Chemistry, and Biology (IFM), Link{\"o}ping University, SE 581 83 Link{\"o}ping, Sweden}
\affil[*]{Corresponding author: thofmann@uncc.edu}

\dates{}

\ociscodes{3D direct laser writing, anti-reflection coating, infrared transmission, infrared reflection}

\doi{}

\begin{abstract}
Three-dimensional direct laser writing via two photon polymerization is used to fabricate anti-reflective structured surfaces composed of sub-wavelength conicoid features optimized to operate over a wide bandwidth in the near-infrared range from 3700~cm$^{-1}$ to 6600~cm$^{-1}$ (2.7 to 1.52~$\mu$m). Analytic Bruggemann effective medium calculations are used to predict nominal geometric parameters such as the fill factor of the constitutive conicoid features of the anti-reflective structured surfaces presented here. 
The performance of the anti-reflective structured surfaces was investigated experimentally using infrared transmission measurements. An enhancement of the transmittance by 1.35\% to 2.14\% over a broadband spectral range from 3700~cm$ ^{-1} $ to 6600~cm$^{-1}$ (2.7 to 1.52~$\mu$m) was achieved. We further report on finite-element-based reflection and transmission data using three-dimensional model geometries for comparison. A good agreement between experimental results and the finite-element-based numerical analysis is observed once as-fabricated deviations from the nominal conicoid forms are included in the model. Three-dimensional direct laser writing is demonstrated here as an efficient method for the fabrication and optimization of anti-reflective structured surfaces designed for the infrared spectral range.
\end{abstract}

\setboolean{displaycopyright}{true}

\begin{document}

\maketitle
\thispagestyle{fancy}

Since the discovery by C.G.~Bernhard in 1967, biomimetic anti-reflective structured surfaces (ARSS) have enjoyed continued interest by the research community \cite{Bernhard1967structural}.
Compared to conventional thin-film anti-reflective coatings, ARSS have some inherent advantages such as the ability to tailor arbitrary index profiles \cite{southwell1983gradient}, to enable frequency-independent Fresnel reflection reduction \cite{stavenga2006light}, or to require only a single material in the fabrication \cite{kowalczyk2014microstructured}, for instance. Theoretically, sub-wavelength features can be designed to completely eliminate Fresnel loss by inducing forward scattering of otherwise evanescent diffraction orders \cite{gaylord1986zero}.

As early as 1973, the functionality of ARSS was established in the visible spectral range \cite{clapham1973reduction}. Using photo-polymerized subwavelength-sized structures fabricated by interference lithography Clapham and Hutley achieved a reduction of the reflectance to less than 0.2\% under normal incidence in the spectral range from 400~nm to 700~nm \cite{clapham1973reduction}. 
In general, complex fabrication techniques have been required for the synthesis of the subwavelength-sized structures which constitute the ARSS and have thereby limited the widespread use of the ARSS approach \cite{lalanne1997antireflection,kanamori1999broadband,yu2003fabrication,kennedy2003porous,xie2008fabrication,ting2009subwavelength}.

Some renewed interest in ARSS is due to advances in nano- and micro-structure fabrication techniques which have led to numerous recent investigations devoted to maximize the transmissivity of ARSS by optimizing the geometry of the ARSS constituent features \cite{li2009biomimetic,min2008bioinspired,han2016antireflective,weiblen2016optimized,kasugai2006light}.

Recently, Kowalczyk \it et al. \rm demonstrated for the first time that 3-dimensional direct laser writing (3D-DLW) via two-photon polymerization can be used for the fabrication of the  subwavelength-sized structures for ARSS in the near-infrared spectral range \cite{kowalczyk2014microstructured}.
This new approach circumvents the restrictions of previously employed lithographic techniques which are only allowed for a limited and very time-consuming exploration of the geometric parameter space of the subwavelength-sized structures. 3D-DLW now allows the fabrication of virtually arbitrary 3D structures with nm-sized features \cite{MarichySR6_2016}. 
As such, 3D-DLW is ideally suited for the fabrication of ARSS and thereby enables optimization of the subwavelength-sized structure geometries required for ARSS in the infrared spectral range \cite{zhang2009morphology,li2010bioinspired}.

Here we demonstrate the application of 3D-DLW for the fabrication of 3D subwavelength-sized structures for broadband, near-infrared ARSS and explore the geometric parameter space of these structures. Our observations show that 3D-DLW allows for rapid fabrication and cost-effective optimization of ARSS designed for the infrared spectral range when compared to previously demonstrated methods.   

Although it is well known that the performance of a given ARSS depends critically on the geometric paramaters of the constitutive features, such as shape, density, height and arrangement, suitable avenues to optimize ARSS performance, however, have not been reported yet. Even though several studies have been carried out to numerically optimize the geometrical parameters \cite{weiblen2016optimized}, only a few results have been experimentally validated. For the near-infrared spectral range, where sub-micron spatial resolution is sufficient to design effective ARSS, 3D-DLW may offer an efficient path for rapid prototyping. Although ARSS performance has been observed over a narrow spectral range centered at 1550~nm, anti-reflection behavior over a broadband spectral range in the near-infrared has not been reported so far \cite{kowalczyk2014microstructured}. 

\begin{figure}[hbt]
	\centering
	\vspace{-0.35cm}
	\includegraphics[keepaspectratio=true, trim=50 60 100 0, clip, width=0.85\linewidth]{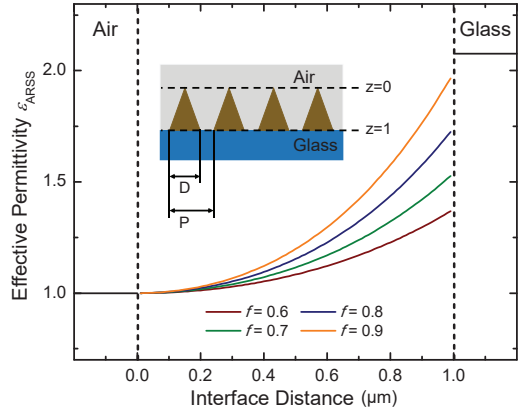}
	\caption{Plot of the real part of the effective permittivity $\varepsilon_{\stext{ARSS}}$ as a function of the distance in micrometers from the top of the ARSS relative to the substrate, $z$, from air ($z=0\ \mu$m) to the substrate ($z=1\ \mu$m) across the ARSS ($0\ \mu$m$<z<$1$\ \mu$m) based on a Bruggemann effective medium homogenization. The effective permittivity at 5000~cm$ ^{-1} $ for cone-like ARSS with $f=$, of 0.6, 0.7, 0.8, and 0.9 are shown as brown, green, blue, and orange solid lines, respectively.  The effective permittivity of air and glass substrate are shown as black solid lines in each corresponding sections. The insert shows the layout of the three individual sections: air, ARSS, and glass substrate. }
	\label{fig:EMAIndex}
\end{figure} 

In this letter, we report on a simple 3D-DLW-based rapid prototyping approach for design and fabrication of conicoid constituent ARSS on transparent substrates to achieve optimal anti-reflective performance over a broadband near-infrared range. Using the Bruggemann effective medium homogenization, we explore reflection reduction and transmission enhancement associated with the variation of the fill factor $f$, defined as the ratio of the cone base diameter $D$ to the lateral separation (pitch) $P$. We find by both experimental FTIR reflection and transmission measurements and three-dimensional finite-element method (3D-FEM) calculations using COMSOL (RF-module) that a maximum increase in transmission of 2.14\% over a spectral range from 3700~cm$^{-1}$ to 6600~cm$^{-1}$ can be achieved for $f=0.8$ for the conicoid structures fabricated here. Scanning electron microcopy (SEM) was performed to evaluate the resulting 3D-DLW geometry of the ARSS constituent features and inform the 3D-FEM calculations.

The ARSS constituents studied here are conicoid sub-wavelength structures symmetrically arranged in a hexagonal lattice pattern. Due to the spatially symmetrical distribution, the Bruggemann effective medium homogenization is suitable to estimate the gradual index transition throughout the ARSS domain \cite{lucarini2005}. The Bruggemann effective medium model for the permittivity of the ARSS $\varepsilon_{\stext{ARSS}}$ accounts for the relative volume fraction of the sub-wavelength structures \cite{Shalaev2009}:
\begin{eqnarray}
\varepsilon_{\stext{ARSS}}(f_{\stext{i}})&=&\frac{1}{4}\Big\{(3f_{\stext{i}}-1)\varepsilon_{\stext{i}}+(2-3f_{\stext{i}})\varepsilon_{\stext{h}}\pm \\\nonumber
&&\sqrt{[(3f_{\stext{i}}-1)\varepsilon_{\stext{i}}+(2-3f_{\stext{i}})\varepsilon_{\stext{h}}]^2+8\varepsilon_{\stext{i}}\varepsilon_{\stext{h}}}\Big\},
\end{eqnarray}
\noindent where $f_{\stext{i}}(f,z)$ is the volume fraction of conicoid subwavelength inclusions $f_{\stext{i}}(f,z)=\frac{\pi f^2(1-z)^2}{2\sqrt{3}}$, and $z$ corresponds to the distance from the top of the ARSS to the substrate. The permittivities of the inclusions (IP-dip) and the host (air) are denoted by $\varepsilon_{\stext{i}}$ and $\varepsilon_{\stext{h}}$, respectively. 

Figure \ref{fig:EMAIndex} depicts $\varepsilon_{\stext{ARSS}}$ at $\omega=\ $ 5000~cm$^{-1}$ as a function of the distance to the air/ARSS interface $z$ for $f=$0.6, 0.7, 0.8, and 0.9. Upon inspection, the smoothest transition of $\varepsilon_{\stext{ARSS}}$ from $\varepsilon_{\stext{h}}$ (air) to the permittivity of the substrate (glass) is found for $f=0.9$. In contrast, the values of $\varepsilon_{\stext{ARSS}}$ for $f=$ 0.6, 0.7, and 0.8 show a stronger permittivity contrast at the ARSS/substrate interface ($z$=1~$\mu$m). 
 
For the purposes of the plotted analytic function in Fig.~\ref{fig:EMAIndex} the domain of the ARSS has been sliced into 100 equally thick sub-layers. The permittivity values for the glass substrate and IP-dip were acquired through variable angle spectroscopic ellipsometry measurements \cite{FullagerOME7_2017}. The cubic profile of the index transition results from the conical shape of the ARSS constituents. 
 
Four ARSS structures composed of conicoid constituents were fabricated by polymerizing IP-dip monomer on glass substrates using a commercially available 3D-DLW system (Photonic Professional GT, Nanoscribe, GmbH). The nominal height of the ARSS is 1~$\mu$m, with the constituent features arranged in hexagonally ordered 50~$\mu$m$\times$50~$\mu$m arrays. The nominal fill factor $f$ is varied from 0.6 to 0.9 in steps of 0.1. The ARSS were fabricated in a single 3D-DLW fabrication step using an optimized exposure dose determined prior to the sample synthesis. After fabrication, the unpolymerized monomer is removed by immersion in PGMEA (Baker 220), and subsequently in 99.99\% isopropyl alcohol for 20 min and 2 min, respectively. Lastly, the remaining isopropyl alcohol is evaporated by blow-drying with dry nitrogen. The samples were post-cured in an ultraviolet oven for ten minutes to ensure complete polymerization of the structures. 

\begin{figure}[hbt]
	\centering
	\includegraphics[keepaspectratio=true, trim=150 190 150 120, clip, width=1\linewidth]{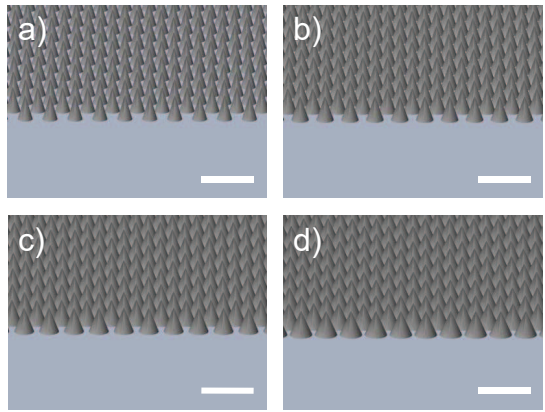}
	\caption{Nominal design for 3D arrays of the conicoid ARSS with a height of 1~$\mu$m and $f=$ 0.6, 0.7, 0.8, and 0.9 shown in panels a), b), c), and d), respectively. Scale bars indicate 2~$\mu$m. }
	\label{fig:nominal}
\end{figure}

\begin{figure}[hbt]
	\centering
	\includegraphics[keepaspectratio=true, trim=150 190 150 120, clip, width=1\linewidth]{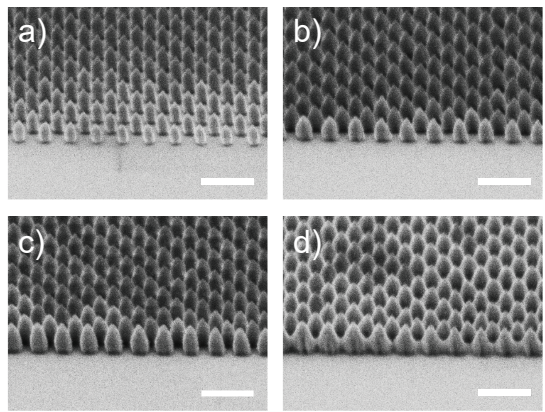}
	\caption{SEM micrographs of the conicoid ARSS with a nominal height of 1~$\mu$m and $f=$ 0.6, 0.7, 0.8, and 0.9 are shown in panels a), b), c), and d), respectively. Note that a SEM micrograph was obtained with the sample tilted by 35$^{\circ}$ from normal. Scale bars indicate  2~$\mu$m. }
	\label{fig:SEM}
\end{figure}

\begin{figure}[hbt]
	\centering
	\includegraphics[keepaspectratio=true, trim=150 190 150 120, clip, width=1\linewidth]{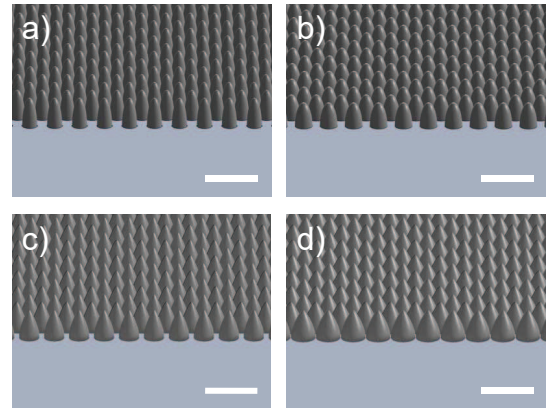}
	\caption{3D arrays of the conicoid ARSS with corrected geometries representative of the as-printed structures for post-comparison with $f=$ 0.6, 0.7, 0.8, and 0.9 shown in panels a), b), c), and d), respectively. Scale bars indicate 2~$\mu$m. }
	\label{fig:print}
\end{figure}

\begin{figure}[hbt]
	\centering
	\includegraphics[keepaspectratio=true, trim=80 45 80 50, clip, width=0.95\linewidth]{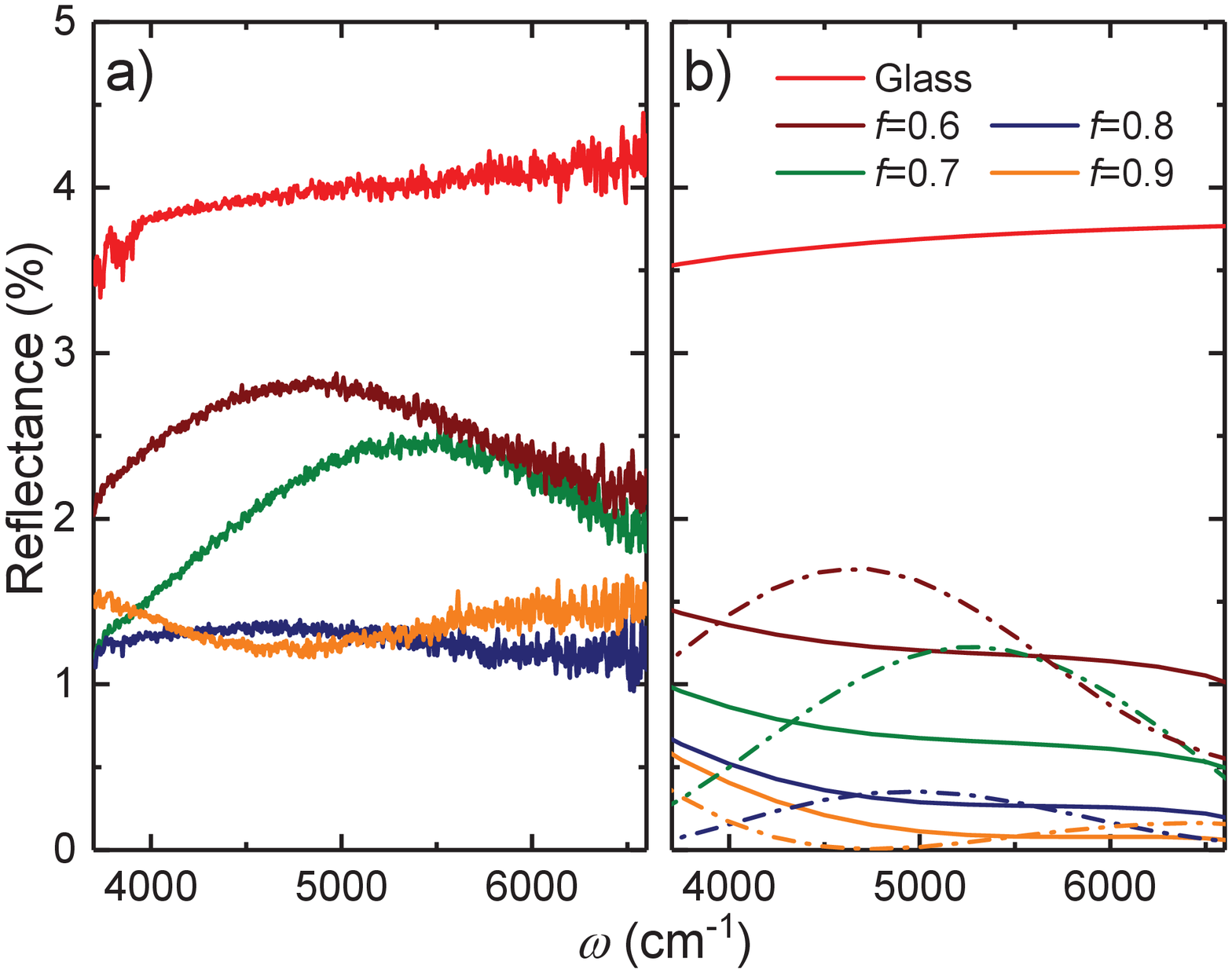}
	\caption{Experimental (a) and 3D-FEM calculated (b) reflectance spectra for $f=$ 0.6 (brown), 0.7 (green), 0.8 (blue), and 0.9 (orange). The solid lines represent the calculated data using the nominal conical geometry, whereas the dash dotted lines indicate the approximated as-printed (see Figs.~\ref{fig:SEM} and \ref{fig:print}) more sharply tapered geometries which were re-iterated in COMSOL.}
	\label{fig:refl}
\end{figure}

3D-FEM calculations were performed for comparison with measured FTIR reflection and transmission data as well as qualitative comparison with the aforementioned homogenized Bruggemann calculations. Representative images depicting the array of nominal conicoid ARSS employed in the 3D-FEM calculations are shown in Fig.~\ref{fig:nominal}. Note that, single unit cells utilizing Floquet (periodic) boundary conditions were used for the 3D-FEM calculations of the ARSS reflectance and transmittance. In addition to the nominal conical geometries, conicoids which more closely resemble the SEM images (see Fig.~\ref{fig:SEM}) were also employed (see Fig.~\ref{fig:print}) as geometrical parameters for the 3D-FEM calculations (Fig.~\ref{fig:refl} b), dashed dotted lines). The dielectric functions of the glass substrate and IP-dip were determined using variable angle IR ellipsometry and are reported elsewhere \cite{FullagerOME7_2017}. An incident angle of 17$^{\circ}$ was used in the numerical finite-element model calculations corresponding to the angular incidence average of the Cassegrain objective used in the infrared microscope \cite{HinrichsJPCC117_2013}.

Figure~\ref{fig:SEM} shows SEM micrographs of the fabricated samples, where panels a), b), c), and d) depict the micrographs of the samples with a nominal fill factor $f$ of 0.6, 0.7, 0.8 and 0.9, respectively. A good agreement between the nominal, as-designed shape and the resulting outcome of the 3D-DLW process is observed for  $f$ of 0.6, 0.7, and 0.8.  As a result of process variability, the $f=$0.9 demonstrates slight deviation from the intended spatial density. 

\begin{figure}[hbt]
	\centering
	\includegraphics[keepaspectratio=true, trim=80 55 80 50, clip, width=0.95\linewidth]{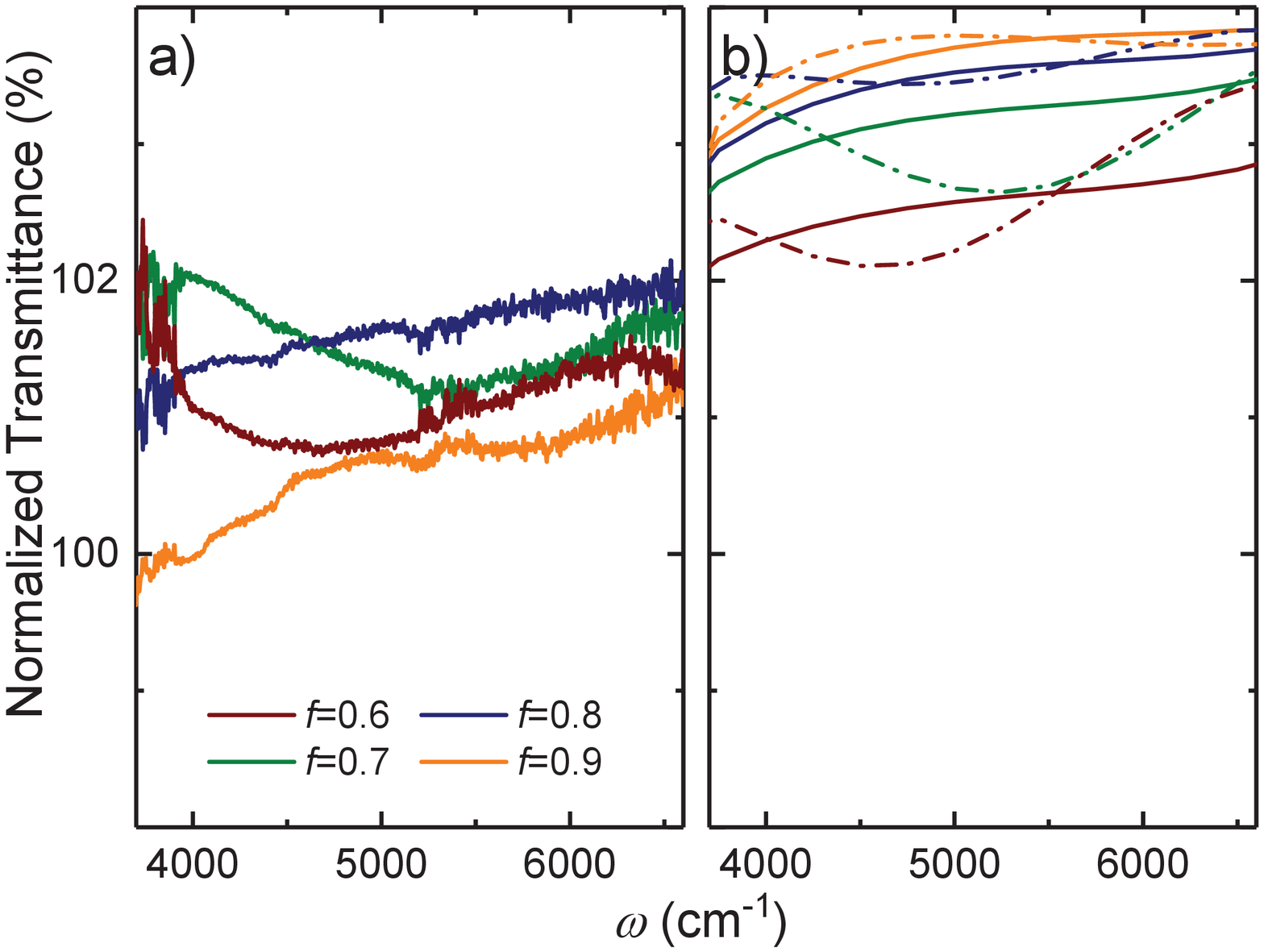}
	\caption{Same as Fig.~\ref{fig:refl} for the experimental and 3D-FEM calculated transmittance data depicted in panels a) and b) respectively. Note that both the experimental and 3D-FEM calculated transmission data were normalized to data obtained experimentally as well as by 3D-FEM calculations from bare glass substrate, respectively.}
	\label{fig:transm}
\end{figure}

Reflection and transmission infrared microscopy measurements were carried out in the spectral range from 3700~cm$ ^{-1} $ to 6600~cm$ ^{-1} $ with a resolution of 4~cm$ ^{-1} $ using a Fourier transform infrared (FTIR) spectrometer in conjunction with an infrared microscope (VERTEX 70 and HYPERION 3000, Bruker, Inc.) and are shown in Figures~\ref{fig:refl} and \ref{fig:transm}, respectively. A 15\texttimes~Cassegrain objective is used for the reflection measurements whereas the transmission measurements used a complementary 15\texttimes~Cassegrain condensor. The Cassegrain illumination configuration results in a range of angles of incidence from 10.8$^{\circ}$ to 23.5$^{\circ}$ defined by the numerical aperture of the Cassegrain objectives and the diameter of their central mirrors \cite{HinrichsJPCC117_2013}. The samples were mounted such that the ARSS faced the objective as opposed to the condensor.

Figure~\ref{fig:refl} shows the comparison of the experimental a) and 3D-FEM model calculated b) reflectance spectra of samples with $f=0.6$ (brown), $f=0.7$ (green), $f=0.8$ (blue) and $f=0.9$ (orange), respectively. The reflectance spectrum of the bare substrate (red solid line) is shown for comparison. The solid and dash dot lines in panel b) represent the calculated reflectance for nominal and as-printed geometries, respectively (see Figs.~\ref{fig:nominal} and \ref{fig:print}). In general, the experimental and 3D-FEM model data are in a good agreement. For all ARSS samples, we observe lower reflectance than for the bare glass substrate. Moreover, with increasing fill factor $f$, the ARSS efficiency increases, corresponding to the trend seen in the 3D model calculation. 
Deviations between the experimental and 3D-FEM model calculated line shapes are attributed to the difference between nominal and as-printed geometries. The corrected 3D-FEM model calculations were reiterated using geometries which more closely resemble the as-printed geometry as observed in the SEM micrographs shown in Fig.~\ref{fig:SEM}. These results are shown using dash dot lines and, in general, show a closer resemblance to the experimental lineshapes. The discrepancy between the experimental and calculated reflectance on average varies by approximately 1\%. 

Figure \ref{fig:transm} shows the comparison of normalized transmittance between the experimental data a) and 3D-FEM model calculation b). The transmission data were normalized to the bare glass substrate. The solid (nominal geometries, see Fig.~\ref{fig:nominal}) and dash dot lines (as-printed geometries, see Fig.~\ref{fig:print}) follow the same trend as a function of $f$ as the reflectance in Fig.~\ref{fig:refl}. The bowing behavior observed in the experimental data for $f=0.6$ and 0.7 is well represented in the 3D-FEM calculations using the as-printed geometries. It is worth noting that the transmittance of the sample with $f=0.9$ has the lowest value which is in contrast to the reflectance spectra in Fig.~\ref{fig:refl}. We attribute this behavior to the structural deviation from the nominal geometry which is most apparent for $f=0.9$ (see Fig.~\ref{fig:SEM}).

In conclusion, we observe that the Bruggemann homogenization detailed herein provides a good analytic prediction of ARSS performance. All the ARSS shown herein exhibit higher transmittance and lower reflectance while being applied to only a single surface than for the bare substrate alone.  Variations in the functionality of the ARSS relative to the 3D-FEM data were investigated through re-simulation utilizing geometrical parameters determined by inspection of SEM micrographs of the ARSS samples. The results obtained in these corrected 3D-FEM calculations showed a better agreement with the experimental reflectance and transmittance data in the cases where there was an apparent discrepancy between the nominal and as-printed conicoids seen in Fig.~\ref{fig:SEM}. Thus, we have demonstrated functional ARSS geometries fabricated using 3D-DLW which operate over a broad bandwidth in the near-infrared spectral range and exhibit negligible absorption. 3D-DLW therefore provides a new avenue for the fabrication and efficient optimization of ARSS for the near-infrared spectral range and might allow for a comprehensive exploration of the parameter space of the ARSS.

\section*{Funding Information}
The authors are grateful for support from the National Science Foundation (1624572) within the I/UCRC Center for Metamaterials, the Swedish Agency for Innovation Systems (2014-04712), and Department of Physics and Optical Science of the University of North Carolina at Charlotte.



\end{document}